\documentclass[%
 reprint,%
 amssymb, amsmath,%
 aps,prc,%
superscriptaddress,
nofootinbib
]{revtex4-1}
\usepackage[colorlinks=true,linkcolor=blue]{hyperref}%
\usepackage{physics}
\usepackage{graphicx}
\usepackage{xcolor}
\usepackage{ulem}

\definecolor{dark-red}{rgb}{0.,0.,0}
\definecolor{dark-blue}{rgb}{0.,0.,1}
\definecolor{medium-blue}{rgb}{0,0,1}
\definecolor{gray}{rgb}{0.85,0.85,0.85}

\hypersetup{
    colorlinks, linkcolor={dark-red},
    citecolor={dark-blue}, urlcolor={medium-blue}
}

\begin{document}

\title{Evolution of $\beta$-decay half-lives in stellar environments}%

\author{A. Ravli\'c }%
\email{aravlic@phy.hr}
\affiliation{Department of Physics, Faculty of Science, University of Zagreb, Bijeni\v{c}ka c. 32, 10000 Zagreb, Croatia}%
\author{E. Y\"uksel}
\email{eyuksel@yildiz.edu.tr}
\affiliation{Physics Department, Faculty of Science and Letters, Yildiz Technical University, Davutpasa Campus, TR-34220 Esenler, Istanbul, Turkey}
\author{Y.F. Niu}
\affiliation{School of Nuclear Science and Technology, Lanzhou University, Lanzhou, China}
\author{N. Paar}
\email{npaar@phy.hr}
\affiliation{Department of Physics, Faculty of Science, University of Zagreb, Bijeni\v{c}ka c. 32, 10000 Zagreb, Croatia}%

\date{\today}%

\begin{abstract}
$\beta$-decay properties of nuclei are investigated within the relativistic nuclear energy density functional framework by varying the temperature and density, conditions relevant to the final stages of stellar evolution. Both thermal and nuclear pairing effects are taken into account in the description of nuclear properties and in the finite temperature proton-neutron relativistic quasiparticle random-phase approximation (FT-PNRQRPA) to calculate the relevant allowed and first-forbidden transitions in the $\beta$-decay. The temperature and density effects are studied on the $\beta$-decay half-lives between temperatures $T = 0-1.5$ MeV, and at densities $\rho Y_e = 10^7$ g/cm${}^3$ and $10^9$ g/cm${}^3$. The relevant Gamow-Teller transitions are also investigated for Ti, Fe, Cd, and Sn isotopic chains at finite temperatures. We find that the $\beta$-decay half-lives increase with increasing density $\rho Y_e$, whereas half-lives generally decrease with increasing temperature. It is shown that the temperature effects decrease the half-lives considerably in nuclei with longer half-lives at zero temperature, while only slight changes for nuclei with short half-lives are obtained. We also show the importance of including the de-excitation transitions in the calculation of the $\beta$-decay half-lives at finite temperatures. Comparing the FT-PNQRPA results with the shell-model calculations for $pf-$shell nuclei, a reasonable agreement is obtained for the temperature dependence of $\beta$-decay rates. Finally, large-scale calculations of $\beta$-decay half-lives are performed at temperatures $T_9(\text{K}) = 5$ and $T_9(\text{K}) = 10$ and densities $\rho Y_e = 10^7$ g/cm${}^3$ and $10^9$ g/cm${}^3$ for even-even nuclei in the range $8 \leq Z \leq 82$, relevant for astrophysical nucleosynthesis mechanisms.
\end{abstract}

\maketitle

\section{Introduction}

Nuclear $\beta$-decay is a fundamental process in atomic nuclei, which plays a decisive role in nuclear astrophysics \cite{r-process-beta-decay-importance, RevModPhys.75.819, MUMPOWER201686} and particle physics \cite{Towner_2010,PhysRev.105.1413,PhysRevC.79.064316} as well as for the properties and structure of nuclei \cite{RevModPhys.75.1021, PhysRevLett.92.232501, PhysRevC.93.014308}. 
Within the context of nuclear astrophysics, recent studies are mainly focused on the understanding of the synthesis of elements heavier than iron via the rapid-neutron capture process ($r-$process) \cite{Burbidge1957547, COWAN1991267, Freiburghaus_1999, 1992ApJ...399..656M}. Along with other nuclear properties (masses, separation energies, etc.), $\beta$-decay half-lives are essential ingredients of the $r-$process calculations, determining the time-scale of the process and the relative abundances of the nuclear species \cite{r-process-beta-decay-importance, RevModPhys.75.819}. It is also known that uncertainties in $\beta$-decay rates can produce significant alterations in the abundance distribution of nuclei \cite{MUMPOWER201686}. Therefore, accurate calculations of $\beta$-decay properties are of utmost importance for $r-$process simulations. Since it is still not possible to reach experimental data for the $\beta$-decay half-lives of all relevant $r-$process nuclei, the simulations mainly rely on theoretical predictions.

The first tabulation of weak interaction rates for stellar environments was done by Fuller, Fowler, Newmann (FFN) \cite{fuller1980stellar,fuller1982stellar_part2,  fuller1982stellar_part3, fuller1985stellar}. Using the independent particle model, rate tables were created for a broad range of temperatures and stellar densities. The shell-model (SM) calculations were also performed to study $\beta$-decay rates of the $sd-$shell nuclei \cite{ODA1994231}, and then extended to the $pf-$shell nuclei \cite{LANGANKE2000481,LANGANKE20011}. Although significant progress has been accomplished over the years, the calculations for heavy nuclei are still demanding due to the huge configuration space of the SM calculations \cite{PhysRevC.85.015802,Suzuki_2016, Suzuki_2018}.
 Large-scale calculations were also performed with the quasiparticle random phase approximation (QRPA) on top of the Finite Range Droplet Model (FRDM) \cite{PhysRevC.67.055802}, which are mostly used in $r-$process simulations today. Apart from the microscopic-macroscopic models, self-consistent models based on the non-relativistic and relativistic energy density functionals were also applied to study $\beta$-decay properties of nuclei \cite{PhysRevC.93.014304,PhysRevC.94.055802,Wang_2016} and their impact on $r-$process calculations \cite{PhysRevC.60.014302, NIU2013172}.
Recently, the relativistic Hartree-Bogoliubov model (RHB) plus relativistic QRPA with momentum-dependent meson-nucleon couplings were used to calculate $\beta$-decay half-lives of neutron-rich nuclei in the $Z \approx 28$ and $Z \approx 50$ regions \cite{PhysRevC.75.024304}. Using a relativistic model with momentum-dependent self-energies in the calculations, predictions for the $\beta$-decay rates were improved. Later on, in Ref. \cite{PhysRevC.93.025805}, the same model was employed to perform large-scale calculation of $\beta$-decay half-lives and $\beta$-delayed neutron emission probabilities of 5409 neutron-rich nuclei in the range $8 \leq Z \leq 124$, including both the allowed (GT) and first-forbidden (FF) transitions. In recent years, $\beta$-decay half-lives of nuclei were also studied using the (quasi)particle-vibration coupling techniques to take into account more complex configurations and obtain a better agreement with the experimental data \cite{marketin12, PhysRevC.85.034314, LITVINOVA2014307,PhysRevC.90.054328,PhysRevLett.114.142501,NIU2018325, LITVINOVA2020135134}.

Investigation of the properties of highly-excited (hot) nuclei is one of the interests in the field of nuclear physics, to better understand the behavior of nuclei under extreme conditions. Over the years, many works have been devoted to study the temperature-driven changes in the nuclear properties as well as the collective excitation properties of nuclei \cite{PhysRevC.88.034308, PhysRevLett.121.082501, PhysRevC.101.044305,LITVINOVA2020135134, FABER19835, PhysRevC.62.054610}. We also note that $\beta$ decays in nuclear astrophysics often occur in various hot stellar environments \cite{1981ApJS45389W, JANKA200738}, thus they have to be described by considering finite temperature effects. For instance, $\beta$-decay can compete with the electron capture in certain stages of the core-collapse supernovae evolution, when the temperatures are in the range $T=1-10$ GK and product of stellar density ($\rho$) and electron-to-baryon ratio ($Y_e$) is $\rho Y_e \geq 10^7$ g/cm${}^3$ \cite{Martinez_Pinedo_2000, PhysRevC.53.3139}. At present,  theoretical description is decisive to provide nuclear properties and processes at finite temperature, necessary for astrophysical modeling.  However, temperature effects on $\beta$-decay rates have been scarcely explored up to now. In Ref. \cite{PhysRevC.80.065808} the finite-temperature QRPA (FT-QRPA) was applied on top of the finite-temperature Skyrme-HF + Bardeen-Cooper-Schrieffer (BCS) theory in order to determine $\beta$-decay half-lives of $N = 82$ isotones. It was shown that the temperature effect first leads to a decrease of the $\beta$-decay half-lives, whereas an increase in the half-lives has been obtained for some open-shell nuclei after $T > 0.6$ MeV. However, Ref.~\cite{PhysRevC.80.065808} includes only Gamow-Teller (GT) excitations in the calculation of the total decay rate. Recently, temperature effects were also studied within the finite-temperature relativistic time-blocking approximation, including nucleon-phonon couplings \cite{LITVINOVA2020135134}. It was shown that the $\beta$-decay rate is quite sensitive to the changes in temperature due to its impact on the low-energy region of the spin-isospin excitations.  

At finite-temperature nucleus can be found in excited states, and one has to take into account transitions between individual excited states both in initial and final nucleus. Within shell-model calculations these transitions can be considered explicitly, weighted by appropriate Boltzmann factors. On the other hand, the FT-QRPA being formulated within a statistical ensamble, contains this information in the form of thermal averages. Importance of considering transitions from highly excited initial states with negative transition energy (de-excitations) has been exemplified by Dzhioev et al. in Refs. \cite{PhysRevC.81.015804,PhysRevC.94.015805,PhysRevC.100.025801,PhysRevC.101.025805,dzhioev2009charge} in the framework of thermal QRPA (TQRPA). Shell-model calculations also incorporate de-excitations from highly excited states in the parent nucleus. They are calculated from the low-lying strength of the inverse process (so-called back-resonance transitions) and corrected in excitation energy by assuming Brink hypothesis \cite{LANGANKE2000481,LANGANKE20011}. 

In this work, we present the first study of the evolution of $\beta$-decay half-lives at finite temperature in stellar environment characterized by a fixed density $\rho Y_e$, including large-scale calculation for even-even nuclei, based on the self-consistent finite temperature proton-neutron relativistic QRPA. Both allowed and first-forbidden transitions are included in description of the $\beta$-decay half-lives at zero and finite temperatures. We choose Ti, Fe, Cd and Sn nuclei to demonstrate the different effects of temperature on the spin-isospin excitations and $\beta$-decay half-lives of open and closed-shell nuclei. Our work provides a theoretical framework capable for microscopic description of temperature-dependent $\beta$-decay rates across the nuclide chart.

We establish a theoretical framework for the description of $\beta$-decay based on the relativistic nuclear energy density functional (RNEDF) with momentum-dependent self-energies \cite{PhysRevC.71.064301}. The nucleons are treated as point-particles that exchange isoscalar-scalar $\sigma-$meson, isoscalar-vector $\omega-$meson and isovector-vector $\rho-$meson (see Refs. \cite{RING1996193,GAMBHIR1990132}). In contrast to usual RNEDFs, additional couplings between nucleon and meson fields are present, containing momentum-dependent terms, thus producing momentum-dependent self-energies. Derivative-coupling (DC) models are known to provide a higher value of the effective nucleon mass $m^*$, giving a higher density of the states around the Fermi level, while still having a good agreement with nuclear-matter and finite-nuclei properties \cite{PhysRevC.75.024304}. In our work, we use the D3C${}^*$ parametrization from Ref. \cite{PhysRevC.75.024304}, which is known to produce a good agreement with experimental values of $\beta$-decay half-lives in medium and heavy nuclei. In order to assess the model dependence of the results and compare with the shell-model calculations, we also employ the effective density dependent meson-exchange interaction DD-ME2 in the calculations \cite{PhysRevC.71.024312}.

This paper is organized as follows. In Sec. \ref{sec:theory} we summarize the theoretical framework used in this work. In Sec. \ref{sec:results}, we study the temperature dependence of the $\beta$-decay half-lives in Ti, Fe, Cd, and Sn nuclei. The importance of including the de-excitation transitions in the calculation of the $\beta$-decay half-lives is also analyzed. The results are presented for the temperature evolution of $\beta$-decay half-lives of selected nuclei in the stellar environment and compared to respective shell-model calculations. Large-scale calculation of $\beta$-decay half-lives is also presented for even-even nuclei in the range $8 \leq Z \leq 82$ at selected stellar densities and temperatures. Finally, conclusions and an outlook are given in Sec. \ref{sec:conclusion}.

\section{Theoretical formalism}\label{sec:theory}

In this work, the finite-temperature Hartree-BCS theory (FT-HBCS) is applied to calculate the nuclear properties of nuclei, and spherical symmetry is assumed in the calculations \cite{GOODMAN198130, yuksel2014effect}. In the present work, only isovector pairing (T=1, S=0) contributes to the FT-HBCS calculations and leads to the partial occupation of states. The isovector pairing strength parameters  $G_{n(p)}$ are adjusted according to the five-point mass formula for each nucleus \cite{PhysRevC.96.024303}.

The charge-exchange excitations are calculated using the finite-temperature proton-neutron relativistic quasiparticle random-phase approximation (FT-PNRQRPA) [see Refs. \cite{PhysRevC.96.024303,PhysRevC.101.044305, yksel2019nuclear, SOMMERMANN1983163} for more details]. Both isovector (T = 1, S = 0) and isoscalar pairing (T = 0, S = 1) contribute in the particle-particle ($pp$) residual interaction part of the FT-PNRQRPA \cite{PhysRevC.67.034312, PhysRevC.69.054303}. For the isoscalar pairing, we employ formulation with a short-range repulsive Gaussian combined with a weaker longer-range attractive Gaussian
\begin{equation}\label{eq:isoscalar_pairing}
V_{12} = V_0^{is} \sum \limits_{j = 1}^2 g_j e^{-r_{12}^2/ \mu_j^2} \prod\limits_{S =1, T=0},
\end{equation}
where $\prod\limits_{S =1, T=0}$ denotes projector on T = 0, S = 1 states.  For the ranges we use $\mu_1$ = 1.2 fm, and $\mu_2$ = 0.7 fm, and strengths are set to $g_1 = 1$ and $g_2 =-2$~\cite{PhysRevC.69.054303}.   For the isovector pairing in residual interaction we employ pairing part of the Gogny interaction \cite{PhysRevC.88.034308}. Isoscalar pairing strength $V_0^{is}$ is considered as a free parameter that can be constrained by the Gamow-Teller excitation or $\beta$-decay experimental data. In 
this work, the functional form introduced in Ref. \cite{NIU2013172} is used,
\begin{equation}\label{eq:ansatz}
V_0^{is}=V_{L}+\frac{V_{D}}{1+e^{a+b(N-Z)}},
\end{equation}
with values $V_L = 153.2(137.8)$ MeV, $V_D $= 8.4(48.7) MeV, $a$ = 6.0(98.6), and $b$ = -0.8(-3.1) for D3C${}^*$(DD-ME2) interaction, adjusted to best reproduce all experimentally available half-life data in the range $8 \leq Z \leq 82$. 
 In the particle-hole channel ($ph$) of the PNRQRPA residual interaction only $\rho-$meson and $\pi$-meson terms are present~\cite{PhysRevC.69.054303}. Due to the derivative nature of pion-nucleon coupling,  the zero-range Landau-Migdal term is also included, that accounts for the contact part of the nucleon-nucleon interaction of the form \cite{PhysRevC.69.054303}
 \begin{equation}\label{eq:landau}
 V_{\delta \pi} = g^\prime \left( \frac{f_\pi}{m_\pi}\right)^2 \boldsymbol{\tau}_1 \boldsymbol{\tau}_2 \boldsymbol{\Sigma}_1 \cdot \boldsymbol{\Sigma}_2 \delta(\boldsymbol{r}_1 - \boldsymbol{r}_2),
\end{equation} where for pion-nucleon coupling standard values are used $m_\pi = 138.0 \text{ MeV}$, $f_\pi^2/(4\pi) = 0.08$, and $\boldsymbol{\Sigma} = \begin{pmatrix}
\boldsymbol{\sigma}  & 0 \\
0 & \boldsymbol{\sigma}
\end{pmatrix}$, $\boldsymbol{\sigma}$ being the Pauli spin matrix and $\boldsymbol{\tau}$ isospin operator. Unless otherwise stated, the strength parameter of the Landau-Migdal term is taken as $g^\prime = 0.76(0.55)$ for D3C${}^*$(DD-ME2) interaction, which is adjusted to reproduce the experimental excitation energy of the Gamow-Teller resonance in ${}^{208}$Pb.

For the calculation of the $\beta$-decay half-lives both allowed ($L=0$) and first-forbidden ($L=0,1$) transitions are included. General form of $\beta$-decay rate in stellar conditions is given by \cite{PhysRevC.89.045806}
\begin{equation}\label{eq:beta-decay}
\lambda=\frac{\ln 2}{K} \int_{0}^{p_{0}} p_{e}^{2}\left(W_{0}-W\right)^{2} F(Z, W) C(W) [ 1 - f(W) ] d p_{e},
\end{equation}
where $W$ is electron energy in the units of $m_e c^2$, $m_e$ denotes  the electron mass, and $p_e$ is electron momentum in units of $m_e c$. $W_0$ is the maximal electron energy given by difference of initial and final nuclear mass. The integration is performed up to a maximal electron momentum $p_0$. $F(Z,W)$ is the Fermi function, taking into account distortion of electron wavefunctions \cite{Kolbe_2003}. Maximal electron energy in $\beta$-decay can be approximated as

\begin{equation}\label{eq:constraint}
W_0 \approx \lambda_{np}+\Delta_{np}-E_{Q R P A},
\end{equation}
where $\lambda_{np} = \lambda_n - \lambda_p$ is the difference between neutron and proton chemical potentials, $\Delta_{np} = 1.293$ MeV is the neutron-proton mass difference, and $E_{QRPA}$ is the FT-PNRQRPA eigenvalue for the considered state. $K$ is measured in superallowed $\beta$-decay to be $K =6144 \pm 2$ s \cite{PhysRevC.79.055502}. $C(W)$ is the so-called shape factor. The outgoing electrons follow a Fermi-Dirac distribution
\begin{equation}\label{eq:fdirac}
f(W) = \frac{1}{\exp \left( \frac{W-\mu_e}{kT} \right) +1},
\end{equation}
where the electron chemical potential $\mu_e$ is determined by the inversion of \cite{PhysRevC.83.045807, PhysRevC.89.045806}
\begin{equation}
\rho Y_e = \frac{1}{\pi^2 N_A} \left( \frac{m_e c}{\hbar} \right)^3 \int \limits_0^{\infty} (f_e - f_{e^+}) p_e^2 dp_e,
\end{equation}
where $\rho$ is the baryon density, $Y_e$ is electron-to-baryon ratio, $N_A$ is Avogadro's number, and $f_{e^+}$ denotes Fermi-Dirac distribution of positrons, for which $\mu_{e^+} = -\mu_{e^-}$. For the allowed GT transitions, $C(W)$ is equal to the reduced matrix element of the GT${}^-$ transition
\begin{equation}\label{eq:gt_strength}
B(\text{GT}^-)=g_{A}^{2} \frac{|\left\langle f\left\|\boldsymbol{\sigma} \boldsymbol{\tau_-}\right\| i\right\rangle|^{2}}{\left(2 J_{i}+1\right)},
\end{equation}
where $\boldsymbol{\tau_-}$ is lowering isospin operator, while $J_i$ is angular momentum of the initial state. Axial-vector coupling constant $g_A$ is quenched from free-nucleon value of $g_A = -1.26$ to $g_A = -1.0$ \cite{PhysRevC.79.054323,PhysRevC.93.025805}. Shape factor for the first-forbidden transitions has functional form
\begin{equation}
C(W)=k+k a W+k b / W+k c W^{2}.
\end{equation}
The details for the definitions of $k, ka, kb, kc$ can be found in Refs.\cite{BEHRENS1971111,PhysRevC.93.025805}. Finally, $\beta$-decay rate $\lambda$ is connected to half-lives $T_{1/2}$ via $T_{1/2} = \text{ln(2)}/ \lambda$. 

It is well known that individual low-energy GT states are decisive in the determination of the $\beta$-decay half-lives of nuclei at zero temperature. At finite-temperature transitions between thermally excited states in both parent and daughter nucleus start to play an important role in determining the total decay rate. Apart from usual transitions included within the FT-(R)QRPA, namely transitions between ground-state of parent nucleus to thermally excited states in daughter as well as transitions from thermally excited states in the parent, in stellar environment it is also important to include the so-called de-excitations, i.e. transitions from highly excited states in the parent nucleus, whose transitions are characterized by the negative $Q-$value. The physical finite-temperature strength function of an external field operator $\hat{F}$ is defined as \cite{fetter2012, RING1984261}
\begin{equation}
\tilde{S}  = \sum \limits_{if} p_i |\langle f | \hat{F} | i\rangle|^2 \delta(E- E_f + E_i),
\end{equation}
where $p_i = e^{-\beta E_i} / \sum_j e^{-\beta E_j}$, $\beta = 1/(k_{B}T)$, $E_{i(f)}$ are the energies of initial (final) states. On the other hand, the FT-(R)QRPA response function is defined as
\begin{equation}
R^{RPA}(E) = \sum \limits_\nu \frac{|\langle \nu | \hat{F} | \tilde{0} \rangle|^2}{E - E_\nu + i \eta} - \frac{|\langle \nu | \hat{F}^\dag | \tilde{0} \rangle|^2}{E + E_\nu + i \eta},
\end{equation}
where $\nu$ labels the FT-(R)QRPA eigenvalue $E_\nu$, while $\ket{\tilde{0}}$ is the vacuum of thermal quasiparticles. A small parameter $\eta$ has been included to add finite width around poles of the response function. For the crucial step, it can be shown \cite{evan_transitions} that the physical strength at finite-temperature can be obtained from the FT-(R)QRPA response function as
\begin{equation}\label{eq:phys_equiv}
\tilde{S}(E) = -\frac{1}{\pi} \text{Im} \left[  \frac{R^{RPA}(E)}{1-e^{-\beta (E-\lambda_{np})}}  \right].
\end{equation}
Using the definition of GT external field operator as $\hat{F} = \boldsymbol{\sigma} \boldsymbol{\tau}_-$ the physical strength is
\begin{equation}
\tilde{S}(E) = \frac{1}{1-e^{-\beta (E-\lambda_{np})}} \left[ S^-(E) + S^+(E) \right],
\end{equation}
where $S^-(E) = \sum_\nu |\langle \nu | \hat{F} | \tilde{0} \rangle|\delta(E-E_\nu) $ and $S^+(E) = -\sum_\nu |\langle \nu | \hat{F}^\dag | \tilde{0} \rangle|\delta(E-E_\nu) $. This means that excitations of thermally averaged initial states are described by GT${}^-$ strength function at positive transition energies, while de-excitations are calculated using the GT${}^+$ strength (induced by $\hat{F}^\dag$ operator) at negative transition energies. We note that negative transition energies mean $E < \lambda_{np}$ for charge-changing transitions.
Finally, by inserting the residues of physical strength function $\tilde{S}$ in the expression for the total $\beta$-decay rate [cf. Eq. (\ref{eq:beta-decay})], it can be rewritten as
\begin{equation}\label{eq:lambda_total_db}
\lambda_\beta = \lambda_\beta^- + \lambda_\beta^+,
\end{equation}
where $\lambda_\beta^-$ is the $\beta$-decay rate calculated using the thermally averaged states with positive transition energy, and $\lambda_\beta^+$ represents the contribution of de-excitations.

\section{Results and discussion}\label{sec:results}
In this part, we study the changes in the $\beta$-decay properties of nuclei alongside the Gamow-Teller excitations at zero and finite temperatures. To this aim, the FT-PNRQRPA with D3C* functional is used in the calculations, which is known to provide a good description of the $\beta$-decay properties of nuclei. The effects of increasing the stellar densities on the $\beta$-decay properties of nuclei are also discussed. In the last part, large-scale calculations are performed for even-even nuclei in the range $8 \leq Z \leq 82$ for the selected stellar densities and temperatures.

\subsection{$\beta$-decay properties at zero and finite temperature}\label{sec:resb}

\begin{figure}
\centering
\includegraphics[width=\linewidth]{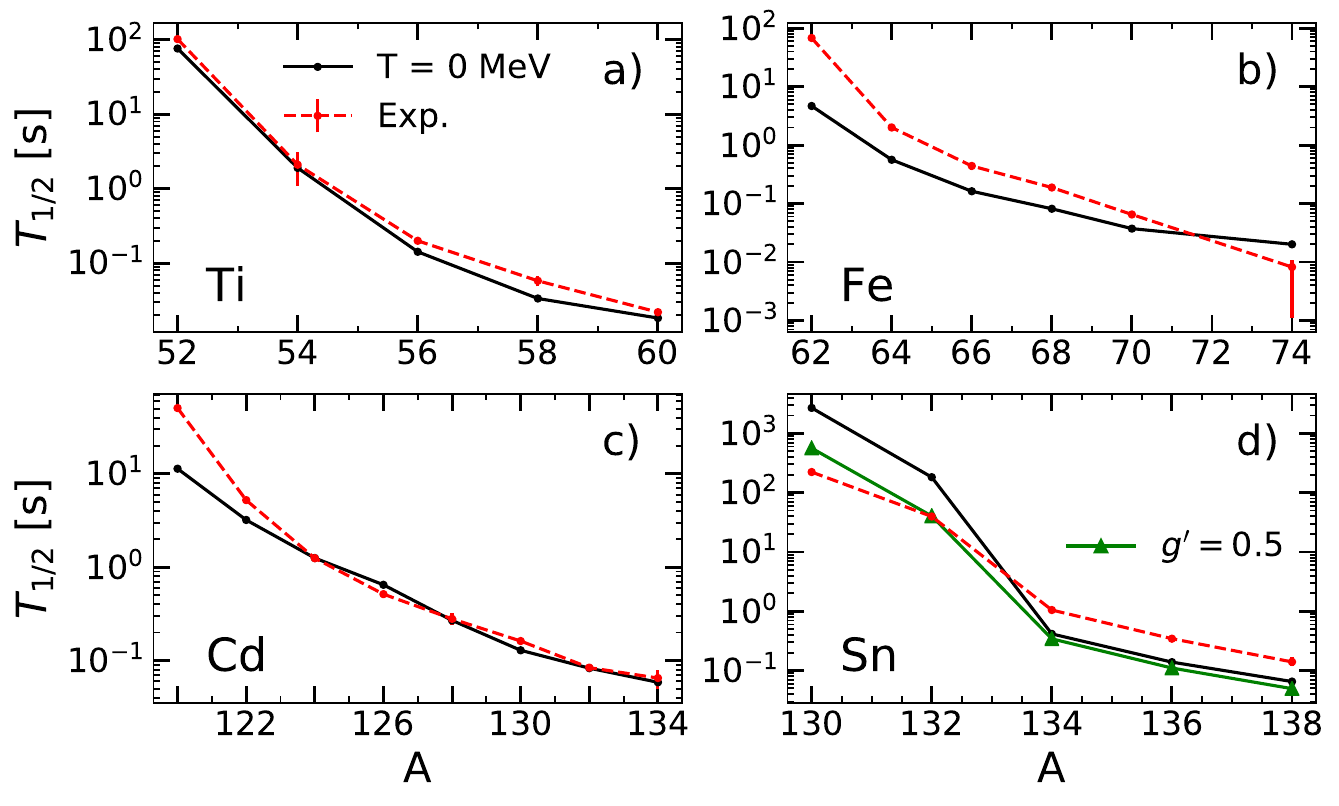}
\caption{Comparison between $\beta$-decay half-lives calculated using the FT-PNRQRPA with the D3C${}^*$ interaction using $g^\prime = 0.76$ (black full line) and the experimental data from Ref. \cite{Audi_2012,nndc} (red dashed line) for Ti, Fe, Cd and Sn isotopic chains. Additionaly, half-lives for the Sn chain obtained by setting $g^\prime = 0.5$ are also shown (green full line). }\label{fig:exp_data}
\end{figure}

As a benchmark for our study, we investigate the $\beta$-decay half-lives in the zero-temperature limit for Ti, Fe, Cd, and Sn isotopes, using D3C* interaction with the Landau-Migdal term strength parameter $g^\prime$=0.76 and the isoscalar pairing strength $V_0^{is}$ as given in Eq. (\ref{eq:ansatz}). The results shown
in Fig. \ref{fig:exp_data} appear in good agreement with the experimental data \cite{Audi_2012,nndc}. In the case of Sn isotopes, additional improvement of the half-lives can be obtained by further adjustment of the $g^\prime$ value, as shown in Fig. \ref{fig:exp_data}(d). It is a known issue that the (Q)RPA consisting of particle-hole (1$p$-1$h$) configuration may overestimate the half-lives of the doubly-magic nuclei \cite{PhysRevC.60.014302, PhysRevC.93.025805}. 
To improve description of the half-lives, contributions due to complex configurations should be taken into account, resulting in lower predictions of half-life by increasing the number of transitions in the low-energy region \cite{PhysRevLett.114.142501, NIU2018325, PhysRevLett.123.202501, LITVINOVA2020135134}. To compensate for the shortcomings of our model for the Sn chain, we induce more strength in the low-energy region by adjusting the strength parameter of the Landau-Migdal term $g^\prime$ to the excitation energy of main  GT${}^-$ peak of ${}^{132}$Sn, and the best fit is obtained for $g^\prime = 0.5$, as shown in Fig. \ref{fig:exp_data}(d). The same value of $g^\prime = 0.5$ is used throughout the whole Sn isotopic chain. The model with the D3C${}^*$ functional benchmarked in this way at T $=0$ MeV, is employed in further studies of the finite temperature effects on $\beta$-decay.

\begin{figure*}
	\centering
	\includegraphics[ width=\linewidth]{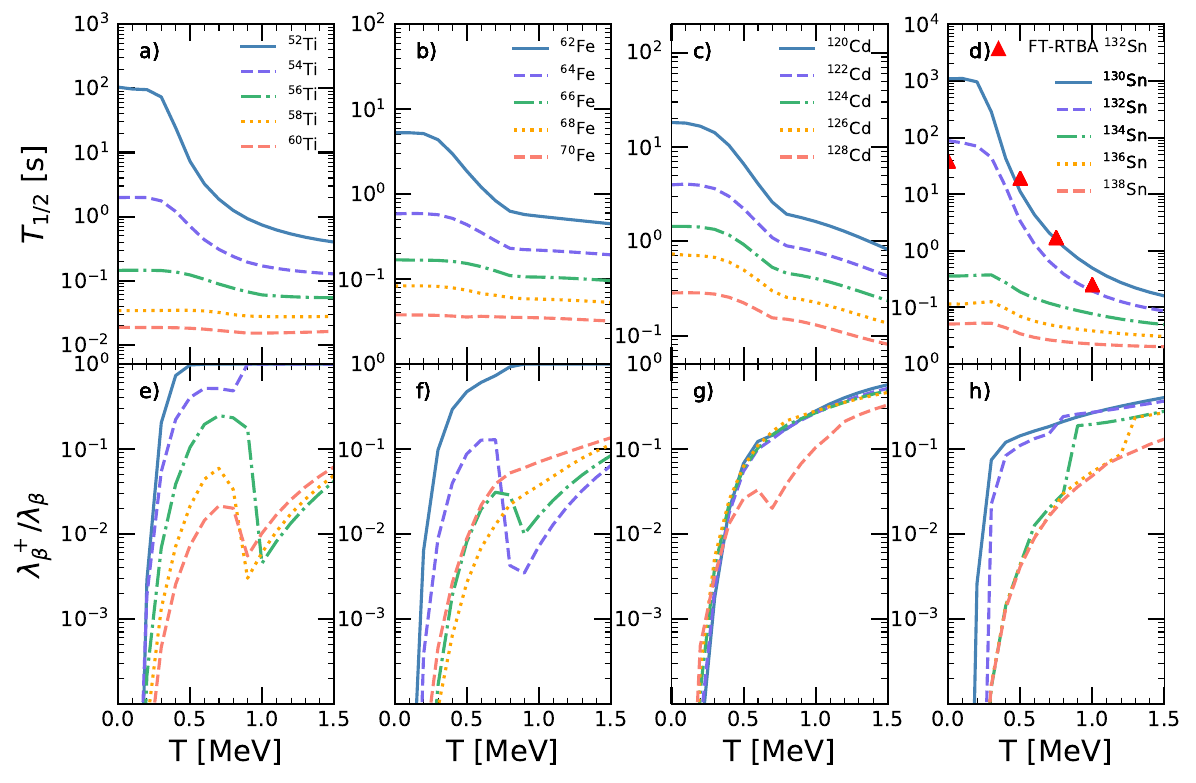}
	\caption{(a)-(d): total $\beta$-decay half-lives $T_{1/2}$ for Ti, Fe, Cd and Sn isotopes as a function of temperature, including negative energy transitions (de-excitations). The calculations are performed using the FT-PNRQRPA and the D3C${}^*$ functional. Red triangles denote results of the FT-RTBA calculation for ${}^{132}$Sn from Ref. \cite{LITVINOVA2020135134}. (e)-(h): ratio between $\beta$-decay rate due to negative energy transitions $\lambda_\beta^+$ and total $\beta$-decay rate $\lambda_\beta$. All calculations are performed at stellar density $\rho Y_e = 10^7$ g/cm${}^3$.}\label{fig_temp_dependence_isotopic_chains}
\end{figure*}

In this work, we first study the temperature evolution of $\beta$-decay half-lives at fixed density ($\rho$) and electron-to-baryon-ratio ($Y_{e}$). Temperature is known to affect both the properties and excitation spectrum of nuclei. Besides, the nucleus goes under a phase transition from a superfluid to a normal state at critical temperatures, and pairing properties vanish completely \cite{SOMMERMANN1983163,PhysRevC.88.034308}. Apart from the changes in the pairing properties and single (quasi)-particle levels of nuclei, temperature also gives rise to the opening of new excitation channels due to the smearing of the Fermi surface and modifies the residual $ph$ and $pp$ interactions of the FT-PNQRPA matrices \cite{PhysRevC.96.024303,PhysRevC.101.044305}. Eventually, the spin-isospin excitations and $\beta$-decay half-lives of nuclei are affected, as we will discuss below.

In figure \ref{fig_temp_dependence_isotopic_chains}(a)-(d), we display the total $\beta$-decay half-lives $T_{1/2}$ of the selected isotopic chains using the FT-PNRQRPA with D3C* functional. Both $\lambda_\beta^-$ and de-excitation $\lambda_\beta^+$ contributions of Eq. (\ref{eq:lambda_total_db}) are taken into account in the calculation of the half-lives, i.e. $T_{1/2} = \text{ln}(2)/(\lambda_\beta^- + \lambda_\beta^+)$. The calculations are performed for the range of temperatures between T $=0-1.5$ MeV and stellar density is fixed to $\rho Y_e = 10^7$ g/cm${}^3$. As a demonstration of the model, we consider Ti, Fe, Cd, and Sn isotopic chains. In Fig. \ref{fig_temp_dependence_isotopic_chains}(a)-(d), it is apparent that temperature has a considerable impact on nuclei with long $\beta$-decay half-lives at zero-temperature, whereas its effect is smaller in short-lived nuclei. For all considered nuclei, the half-lives almost do not change up to T $ \approx 0.3$ MeV, above which they start to decrease or slightly increase and converge to an almost constant value at higher temperatures. As mentioned above, influence of temperature is more pronounced for nuclei with long half-lives at zero-temperature, e.g., ${}^{52}$Ti, ${}^{62}$Fe, ${}^{120}$Cd, and ${}^{130,132}$Sn where we first observe a sharp decrease of half-lives with increasing temperature. As we will discuss in Section \ref{sec:GT}, this sharp decrease in the $\beta$-decay half-lives at low temperatures is related to the changes in the pairing correlations as well as the changes in the low-energy states and contribution of negative energy transitions (de-excitations) in the calculations, which eventually increase the $\beta$-decay phase space, and decrease the half-lives.
It is also seen that nuclei with shortest half-lives at the zero-temperature, like ${}^{60}$Ti and ${}^{70}$Fe, show almost no temperature dependence. The conclusion by inspecting Fig. \ref{fig_temp_dependence_isotopic_chains}(a)-(d) is that in general temperature leads to a decrease of $\beta$-decay half-lives, effect being larger (smaller) for nuclei with longer (shorter) half-lives at zero temperature. In Fig. \ref{fig_temp_dependence_isotopic_chains}(d), we also display the half-life of ${}^{132}$Sn (red triangles) calculated within the FT-RTBA formalism from Ref. \cite{LITVINOVA2020135134} at $\rho Y_e = 10^7$ g/cm${}^3$. The overall trend of the temperature dependence agrees well between the FT-PNRQRPA and FT-RTBA calculations, that is, the half-life decreases with increasing temperature. While our calculation predicts first significant temperature effect at T $ \approx 0.4$ MeV, the FT-RTBA predicts visible effects starting from T $ \approx 0.5$ MeV.
At T $ = 1$ MeV, the results of both approaches agree reasonably well. It should be noted that both the framework of the compared models and the effective nuclear interactions used in the calculations are different, which in turn results in different predictions for the single-particle energies and transitions relevant for the $\beta$-decay. Furthermore, within our model, the chemical potential of electrons at lower-temperatures ($\mu_e \approx 1$ MeV) is large enough to slightly decrease the $\beta$-decay rate and thus increase the half-life. Considering all of these reasons, the differences between the FT-PNRQRPA and FT-RTBA calculations for the half-lives of nuclei can be expected already at zero temperature.

\begin{figure*}
	\centering
	\includegraphics[ width=\linewidth]{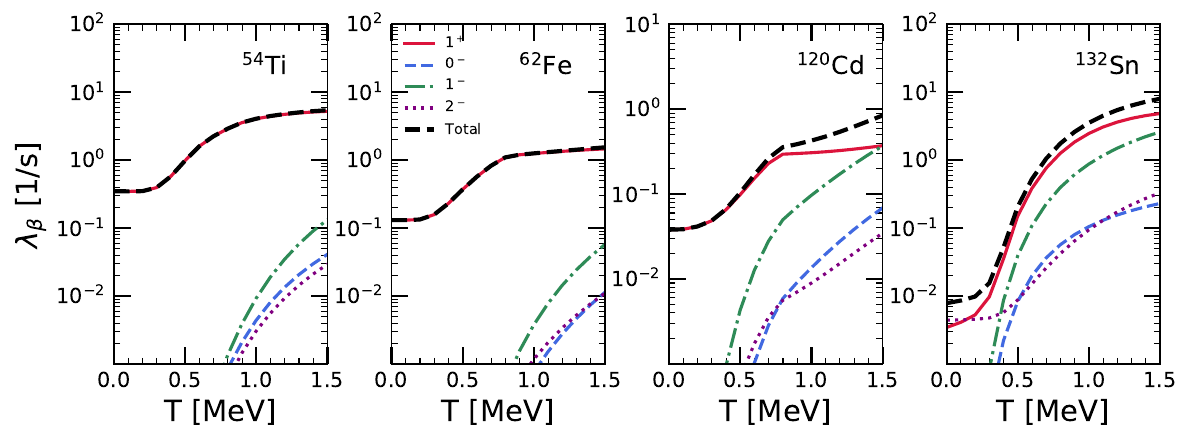}
	\caption{Temperature dependence of $\beta$-decay rates $\lambda_\beta$ for both allowed ($1^+$) and first-forbidden transitions ($0^-, 1^-, 2^-$) together with their total sum (Total) for ${}^{54}$Ti, ${}^{62}$Fe, ${}^{120}$Cd and ${}^{132}$Sn. Calculations are performed at stellar density $\rho Y_e = 10^7$ g/cm${}^3$.}\label{fig_temp_dependence_ff}
\end{figure*}

To demonstrate the importance of including de-excitations in the calculation of $\beta$-decay half-lives, in Fig. \ref{fig_temp_dependence_isotopic_chains}(e)-(h), we also display the temperature evolution of the ratio between de-excitation rate $\lambda_\beta^+$ and total $\beta$-decay rate $\lambda_\beta$ for the same isotopic chains [cf. Eq. (\ref{eq:lambda_total_db})].
It is found that (i) negative energy transitions (de-excitations) start to play a role already at T $\approx$ 0.3 MeV, (ii) its contribution increases for all considered nuclei with increasing temperature, (iii) for $pf-$shell ${}^{52,54}$Ti and ${}^{62}$Fe nuclei, $\beta$-decay rates are fully determined by de-excitation transitions around critical temperatures for pairing correlations (iv) temperature evolution of $\lambda_\beta^+ / \lambda_\beta$ depends on shell-structure of particular nucleus under consideration. For most nuclei in Ti and Fe chain and some in Cd and Sn isotopic chain, a sharp decrease of $\lambda_\beta^+$ contribution occurs in the vicinity of critical temperature for pairing phase transitions. As we will demonstrate later in the section, this occurs due to redistribution of the GT strength function when the pairing correlations vanish at critical temperatures. It is shown that inclusion of the negative energy transitions has the most considerable effect on $pf$-shell nuclei with longer half-lives, and leads to a smooth decrease in the $\beta$-decay half-lives with increasing temperature. Also, the contribution of the negative energy transitions increases with increasing temperature for all nuclei considered in this work.

Apart from the Gamow-Teller states, the first-forbidden transitions are also known to play a crucial role in the determination of the $\beta$-decay half-lives in certain regions of the nuclear chart \cite{PhysRevC.93.025805}. Therefore, in the present analysis, both allowed and first-forbidden transitions are considered for a proper description of half-lives at finite temperatures. In Fig. \ref{fig_temp_dependence_ff}, we display the contribution of the GT ($1^+$ transition) and first-forbidden ($0^-, 1^-, 2^-$) transitions to the total $\beta$-decay rate in ${}^{54}$Ti, ${}^{62}$Fe, ${}^{120}$Cd and ${}^{132}$Sn with increasing temperature. It is shown that for ${}^{54}$Ti and ${}^{62}$Fe the $\beta$-decay rate is dominated by the allowed GT transition up to T = 1.5 MeV. The first-forbidden transitions also start to contribute to the $\beta$-decay rate with increasing temperature, whereas their contribution is quite less and can be neglected up to T $ = 1.5$ MeV. Although their impact on the total $\beta$-decay rate is rather small compared to the GT transitions up to T $\approx$ 0.5 MeV, forbidden transitions start to contribute to the $\beta$-decay rate of neutron-rich ${}^{120}$Cd after T $\approx$ 0.5 MeV. On the other hand, at low temperatures (T $ < 0.3$ MeV), both GT and $2^-$ transitions are important for the $\beta$-decay rate in ${}^{132}$Sn. By increasing temperature further, the contribution of the GT excitations increases considerably, while the contribution of $2^-$ multipole to the total rate cannot compete with GT and becomes less. At T $\approx$ 1.5 MeV, contribution of $1^-$ multipole has a nonnegligible contribution to the $\beta$-decay rates in ${}^{120}$Cd and ${}^{132}$Sn. Our results show that the contribution of the first-forbidden transitions increases with increasing temperature.

Even though one drawback of our model is the inclusion of only two quasiparticle (q.p.) configurations within the R(Q)RPA, it includes both the pairing and temperature effects and can be applied to calculations throughout the nuclide chart, allowing large-scale calculations of relevance for astrophysical models of stellar evolution and synthesis of chemical elements. The deformation is another factor that can affect the $\beta$-decay properties of nuclei. However, it is also known that deformation of nuclei decreases with increasing temperature, and a second-order phase transition from the deformed state to the spherical state generally occurs after T$>$1 MeV \cite{SAXENA2019323, PhysRevC.93.024321, PhysRevC.62.044307}. Therefore, our model applies well for the extreme stellar environments where the temperature is high enough (e.g. core-collapse supernovae).

\subsection{Gamow-Teller excitations at zero and finite temperature}\label{sec:GT}

To understand the temperature evolution of $\beta$-decay half-lives we need to investigate changes in the spin-isospin excitations in nuclei with increasing temperature. To keep the discussion simple, we fix the stellar density to $\rho Y_e = 10^7$ g/cm${}^3$ and consider only the allowed Gamow-Teller transitions, whose strength is defined in Eq. (\ref{eq:gt_strength}). In this section we study the changes in the GT strength of two $pf-$shell nuclei ${}^{54}$Ti and ${}^{62}$Fe with increasing temperature. To start with, the pairing gap values at zero temperature and the critical temperatures $T^c_{n(p)}$ for neutrons (protons) of ${}^{54}$Ti and ${}^{62}$Fe are given in Table \ref{tab:tab_delta}. Due to the grand-canonical treatment of nuclei at finite temperatures, pairing phase transition of nuclei from superfluid state to normal state occurs at critical temperatures, and pairing correlations disappear. Temperature leads to a decrease in the excited state energies of nuclei due to the vanishing of the isovector pairing properties and changes in the single(quasi)-particle energies of nuclei as well as the decrease in the residual $ph$ and $pp$ interactions. Besides, new excitation channels become possible due to the smearing of the Fermi surface at high temperatures \cite{PhysRevC.63.032801, NIU2009315, PhysRevC.96.024303, PhysRevLett.121.082501, LITVINOVA2020135134, PhysRevC.101.044305}. However, presence of the isoscalar pairing in the residual interaction can slow down temperature-induced changes for open-shell nuclei until the critical temperatures. As mentioned before, both the isoscalar and isovector pairing correlations are included in the calculations at zero and finite temperatures. While the isovector pairing contributes to the FT-HBCS calculations and leads to an increase in both the quasiparticle energies of the states and excited state energies, the isoscalar pairing contributes to the residual pairing interaction and decreases the excited state energies due to its attractive nature. By increasing temperature, pairing effects first decrease and then vanish completely at critical temperatures. Under the influence of both the isoscalar and isovector pairing, the changes in the GT${}^-$ excitations strongly depend on the interplay between the increasing effect of temperature and the decreasing impact of the pairing correlations for the considered nucleus (see Ref. \cite{PhysRevC.101.044305} for more details).

\begin{figure}[ht!]
\centering
\includegraphics[width=\linewidth]{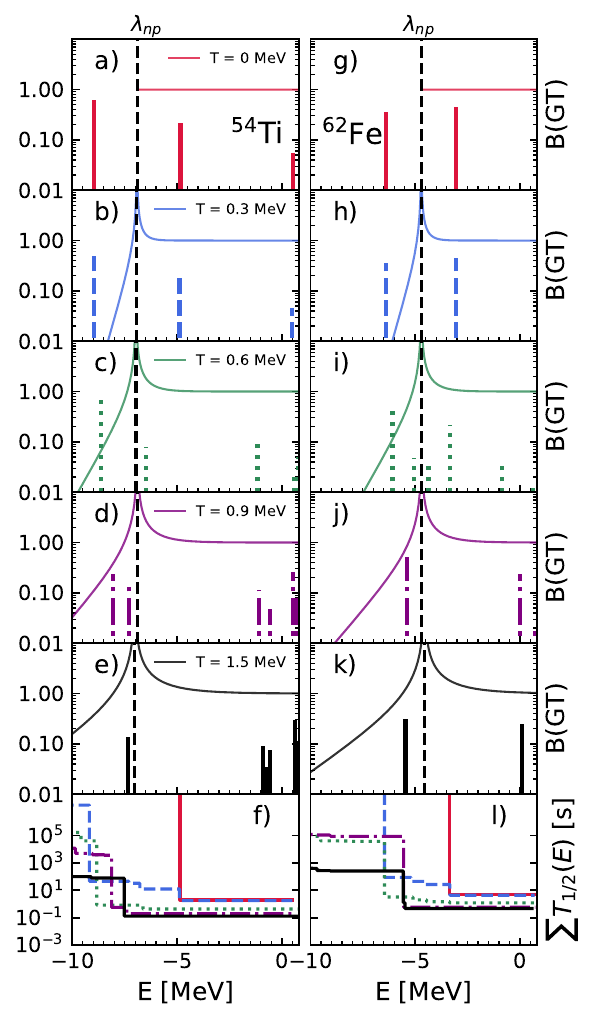}
\caption{The temperature evolution of the Gamow-Teller strength, located within the $Q_\beta$ window, in ${}^{54}$Ti (panels (a)-(e)) and ${}^{62}$Fe (panels (g)-(k)) at temperatures in the range T = 0-1.5 MeV with respect to the excitation energy of parent nucleus. The neutron-proton chemical potential difference $\lambda_{np} = \lambda_n - \lambda_p$ is labeled by the black dashed line and it separates the negative energy transitions (determined by GT${}^+$ strength located at $E < \lambda_{np}$) from GT${}^-$ strength (located at $E > \lambda_{np})$. Within the panels (a)-(e) (and (g)-(k)) thick lines denote the Gamow-Teller strength B(GT) while thin lines represent the weighting prefactors $(1-e^{-\beta (E-\lambda_{np})})^{-1}$. In panels (f) and (l) the cummulative sum of half-lives is shown obtained by restricting the summation in Eq. (\ref{eq:beta-decay}) for ${}^{54}$Ti and ${}^{62}$Fe, respectively.}\label{fig:gtm_half_life}
\end{figure}

In Fig. \ref{fig:gtm_half_life}(a)-(e) we show the temperature evolution of the GT strength in ${}^{54}$Ti within the $\beta$-decay energy window (i.e. $Q_\beta$ window). 
As mentioned above, thermally induced negative energy transitions have a considerable impact on the calculation of the $\beta$-decay half-lives below the critical temperatures. To explain the working mechanism of including de-excitations in our model, we display the finite-temperature GT strength in Fig. \ref{fig:gtm_half_life}(a)-(e). The negative energy strength representing de-excitations is located up to $\lambda_{np}$ (denoted by black dashed vertical line in Fig. \ref{fig:gtm_half_life}(a)-(e)). In the same figure, the positive energy strength is also displayed between $\lambda_{np}$ and $\Delta_{n H}$. At finite-temperature, both positive energy strength, determined by the GT${}^-$ transitions for $E>\lambda_{np}$ and negative energy strength, determined by the GT${}^+$ transitions for $E < \lambda_{np}$, contribute to the decay rate, where we have denoted the corresponding rates with $\lambda_\beta^-$ and $\lambda_\beta^+$ [cf. Sec. \ref{sec:theory}]. Strength functions corresponding to these rates are weighted by the temperature factor $(1-e^{-\beta (E-\lambda_{np})})^{-1}$ (shown as thin colored lines in panels (a)-(e) of Fig. \ref{fig:gtm_half_life}).


\begin{table}
\caption{The pairing gap values at zero temperature $\Delta^0_{n(p)}$ and critical temperatures $T^c_{n(p)}$ for neutrons (protons) calculated using the FT-HBCS theory and D3C${}^*$ functional.}\label{tab:tab_delta}
\begin{tabular}{ccccc}
\hline
\hline
 & $\Delta^0_n$ [MeV] & $\Delta^0_p$ [MeV] & $T^c_n$ [MeV] & $T^c_p$ [MeV] \\
\hline
${}^{54}$Ti & 1.03 & 1.76 & 0.59 & 0.91 \\
${}^{62}$Fe & 1.44 & 1.56 & 0.76 & 0.83 \\
\hline
\end{tabular}
\end{table}

At T $ = 0$ MeV, the main low-energy GT${}^-$  peak within the $Q_\beta$ energy window is found at $E = -4.83$ MeV with strength B(GT${}^-$) = 0.22  (see panel (a) of Fig.\ref{fig:gtm_half_life}). Although this is not the only peak within the $Q_\beta$ window, it is the only one allowed by the phase-space factor in Eq. (\ref{eq:beta-decay}) at $\rho Y_e = 10^7$ g/cm$^3$. Most of the GT${}^-$ strength comes from $\nu 1f_{5/2} \rightarrow \pi 1f_{7/2}$ transition for this state, where $\nu$ and $\pi$ refer to neutron and proton, respectively. At T = 0 MeV, weighting factor $(1-e^{-\beta (E-\lambda_{np})})^{-1}$ reduces to 1, which is multiplied by the GT${}^-$ strength for the calculation of the $\lambda_\beta^-$ rate. Besides, there is a peak for $E < \lambda_{np}$ at E = -8.97 MeV. However, this state does not contribute to the $\beta$-decay rate since the weighting factor is zero for the states below $E < \lambda_{np}$. In Fig. \ref{fig:gtm_half_life}(f), we also display the cumulative sum of the $\beta$-decay half-lives to follow the changes on the GT excitations and $\beta$-decay properties of nuclei with increasing temperature.  At T $ = 0$ MeV, the effect of the main low-energy peak at $E = -4.83$ MeV can be seen clearly in reducing the half-life, while other peaks are being negligible. Already at T = 0.3 MeV, temperature affects the GT strength function. The main peak within the $Q_\beta$ window shifts by 0.05 MeV to lower energies, and is obtained at E = -4.88 MeV. Since the weighting factor is greater than zero for the states below $E < \lambda_{np}$, the peak at E = -8.95 MeV with B(GT${}^+$) = 0.63 also contributes to the $\lambda_\beta^+$ part of the total $\beta$-decay rate. It is a negative energy transition originating from $\pi 1 f_{7/2} \to \nu 1f_{5/2}$ two-quasiparticle (2qp) excitation. The impact of this transition in decreasing the half-life can be seen in Fig. \ref{fig:gtm_half_life}(f). 

At T $ = 0.6$ MeV, pairing collapse occurs for neutron states (cf. Tab. \ref{tab:tab_delta}), and considerably changes the GT strength function as can be seen from panel (c) of Fig. \ref{fig:gtm_half_life}. The most important GT${}^-$ peaks for $E > \lambda_{np}$ are: the peak at $E = -6.48$ MeV with B(GT${}^-$) = 0.08 with main contribution from $\nu 1f_{5/2} \rightarrow \pi 1f_{7/2}$ transition and another low-enegy peak at $E = -1.15$ MeV with B(GT${}^-$) = 0.09 mainly formed with the $\nu 2p_{1/2} \rightarrow \pi 2p_{3/2}$ transition. Besides, the new peaks appear with the opening of new excitation channels, and the low-energy strength fragments at higher energies (E $>$ 0 MeV). These excited states do not play an important role in the $\beta$-decay half-lives as can be seen from Fig. \ref{fig:gtm_half_life}(f). For $E < \lambda_{np}$ the main peak is located at E = - 8.64 MeV with B(GT${}^+$) = 0.70, again stemming from $\pi 1 f_{7/2} \to \nu 1f_{5/2}$ transition. Due to the growing impact of the weighting factor $(1-e^{-\beta (E-\lambda_{np})})^{-1}$ for $E < \lambda_{np}$, this peak gains considerable importance by lowering the half-life as can be seen in Fig. \ref{fig:gtm_half_life}(f). At T $ = 0.9$ MeV the pairing gap of proton states reduces further to $\Delta_p = 0.54$ MeV, while there is no pairing for neutron states (see Table \ref{tab:tab_delta}). The GT${}^-$ peaks for $E > \lambda_{np}$ are found at higher excitation energies. The first important peak is obtained at E = -1.08 MeV with strength B(GT${}^-$) = 0.11, originating from $\nu 2p_{1/2} \to \pi 2p_{3/2}$ transition. Another low-energy peak is found at E = -0.57 MeV with B(GT${}^-$) = 0.05, having contribution from $\nu 1f_{5/2} \to \pi 1f_{5/2}$ and $\nu 1f_{7/2} \to \pi 1f_{7/2}$ transitions.  However, these peaks together with other peaks located at E $>$ 0 MeV have almost no contribution to the $\beta$-decay half-life. At T $=0.9$ MeV, the $\beta$-decay half-lives are almost fully determined by de-excitations, located on the $E < \lambda_{np}$ side. The first peak is obtained at E = -8.04 MeV with B(GT${}^+$) = 0.24 and the second one is found at E = -7.29 MeV with B(GT${}^+$) = 0.15, both stemming from the $\pi 1f_{7/2} \to \nu 1f_{5/2}$ negative energy transition. Combined with the increasing impact of the weighting factor, the inclusion of de-excitations leads to a smooth decrease in the half-lives with increasing temperature (see fig. \ref{fig_temp_dependence_isotopic_chains}(a)). At T = 1.5 MeV, pairing effects are washed out completely, and new transitions appear because of the unblocking effect of the temperature. The fragmentation of the states also increases around E $\approx 0$ MeV. Although the overall strength increases with the opening of the new excitation channels at finite temperatures, the strength mostly lies in the higher energy region of the $Q_\beta$ window and its effect on the half-life is negligible. At T $=1.5$ MeV, the most important contribution to the total $\beta$-decay rate comes from the negative energy transition at E = -7.36 MeV with B(GT${}^+$) = 0.24, which is mainly formed by the $\pi 1f_{7/2} \to \nu 1f_{5/2}$ transition. Notice that at T $= 1.5$ MeV half-life slightly decreases compared to T = 0.9 MeV.

\begin{table*}
\caption{The most dominant Gamow-Teller excitations contributing to $\beta$-decay half-lives of ${}^{62}$Fe in the temperature range T = 0-1.5 MeV. Displayed are both $E > \lambda_{np}$ (determined by GT${}^-$) and $E < \lambda_{np}$ (determined by GT${}^+$) excitations, together with their excitation energy E (w.r.t. parent nucleus), strength B(GT${}^\pm$), absolute value of weighting factor $|(1-e^{-\beta (E-\lambda_{np})})^{-1}|$ and two-quasiparticle transitions having most relevant contributions.}\label{tab:dominant_gt_62fe}
\begin{tabular}{cccccc}
\hline
\hline
T [MeV] & & E [MeV] & B(GT) & $|(1-e^{-\beta (E-\lambda_{np})})^{-1}|$ & transitions \\
\hline
0.0 & GT${}^-$ & -3.04 & 0.45 & 1.00 & $\nu 1f_{5/2} \to \pi 1f_{7/2}$ \\
    & GT${}^+$ & -6.39 & 0.36 & 0.00 & $\pi 1f_{7/2} \to \nu 1f_{5/2}$ \\
0.3 & GT${}^-$ & -3.05 & 0.44 & 1.00 & $\nu 1f_{5/2} \to \pi 1f_{7/2}$ \\
	& GT${}^+$ & -6.37 & 0.37 & 0.004 & $\pi 1f_{7/2} \to \nu 1f_{5/2}$ \\
0.6 & GT${}^-$ & -4.36 & 0.04 & 2.29 & $\nu 1f_{5/2} \to \pi 1f_{7/2}$ \\
    &          &  -3.33 & 0.22 & 1.11 & $\nu 1f_{5/2} \to \pi 1f_{7/2}$ \\
    &          &  -0.84 & 0.04 & 1.00 & $\nu 2p_{1/2} \to \pi 2p_{3/2}$ \\  
	& GT${}^+$ & -5.05 & 0.05 & 1.29 & $\pi 1f_{7/2} \to \nu 1f_{5/2}$ \\
	&          & -6.08 & 0.42 & 0.11 & $\pi 1f_{7/2} \to \nu 1f_{5/2}$ \\
0.9 & GT${}^-$ & 0.02 & 0.24 & 1.01 & $\nu 2p_{1/2} \to \pi 2p_{3/2}$ \\
	& GT${}^+$ & -5.38 & 0.51 & 0.86 & $\pi 1f_{7/2} \to \nu 1f_{5/2}$ \\
1.5 & GT${}^-$ & 0.21 & 0.25 & 1.05 & $\nu 2p_{1/2} \to \pi 2p_{3/2}$ \\
	& GT${}^+$ & -5.36 & 0.31 & 1.22 & $\pi 1f_{7/2} \to \nu 1f_{5/2}$ \\
    &  &  &  &  & $\pi 1g_{9/2} \to \nu 1g_{7/2}$ \\
    &  & -5.38 & 0.18 & 1.17 & $\pi 1f_{7/2} \to \nu 1f_{5/2}$ \\
    &  &  &  &  & $\pi 1g_{9/2} \to \nu 1g_{7/2}$ \\
\hline	
\end{tabular}
\end{table*}

Temperature dependence of GT strength for ${}^{62}$Fe is also shown in Fig. \ref{fig:gtm_half_life}(g)-(k). At zero temperature, only contribution to GT${}^-$ strength comes from $\nu 1f_{5/2} \to \pi 1f_{7/2}$ 2qp excitation and we obtain an excited state at E $=-3.04$ MeV with B(GT${}^-$) = 0.45. By increasing temperature, the negative energy $\pi 1f_{7/2} \to \nu 1f_{5/2}$ transition becomes allowed due to non-vanishing weighting factor multiplying the strength. Similar to the findings in  ${}^{54}$Ti, de-excitations start to have a dominant contribution to the half-life with increasing temperature (see Fig. \ref{fig:gtm_half_life}(l)).  At T = 0.9 MeV pairing correlations vanish for both proton and neutron states (check Tab \ref{tab:tab_delta}). After the pairing collapse, the GT${}^-$ peak is obtained at E $\approx$ 0 MeV, which is formed with the $\nu 2p_{1/2} \to \pi 2 p_{3/2}$ transition and has almost no contribution to $\beta$-decay half-lives. At the same time, the impact of the de-excitations increases which in turn leads to a smooth decrease in the half-lives. At T = 1.5 MeV, $E < \lambda_{np}$ strength is fragmented into two peaks due to temperature unblocking, and both $\pi 1f_{7/2} \to \nu 1f_{5/2}$ and $\pi 1g_{9/2} \to \nu 1g_{7/2}$ transitions contribute to the $\lambda_\beta^+$ rate. In Tab. \ref{tab:dominant_gt_62fe}, we also provided the most important peaks for the $\beta$-decay half-lives alongside their excitation energy E, strength B(GT), absolute value of weighting factor $|(1-e^{-\beta (E-\lambda_{np})})^{-1}|$, and main 2qp components for ${}^{62}$Fe.  Evolution of $\beta$-decay half-lives in Fig. \ref{fig_temp_dependence_isotopic_chains}(a)-(d) for ${}^{54}$Ti and ${}^{62}$Fe precisely follows these trends of GT strength, mainly (i) at low temperatures (T $<$ 0.5 MeV), the rate is determined only by a few GT peaks in $Q_\beta$ window yielding long half-lives, (ii) once the temperature reaches T $\approx$ 0.6 MeV half-lives are dominated by the de-excitations and (iii) due to inclusion of higher number of states in GT strength at high temperatures (T $>$ 1 MeV) rates are less dependent on temperature effects.

\begin{figure*}
\includegraphics[width = \linewidth]{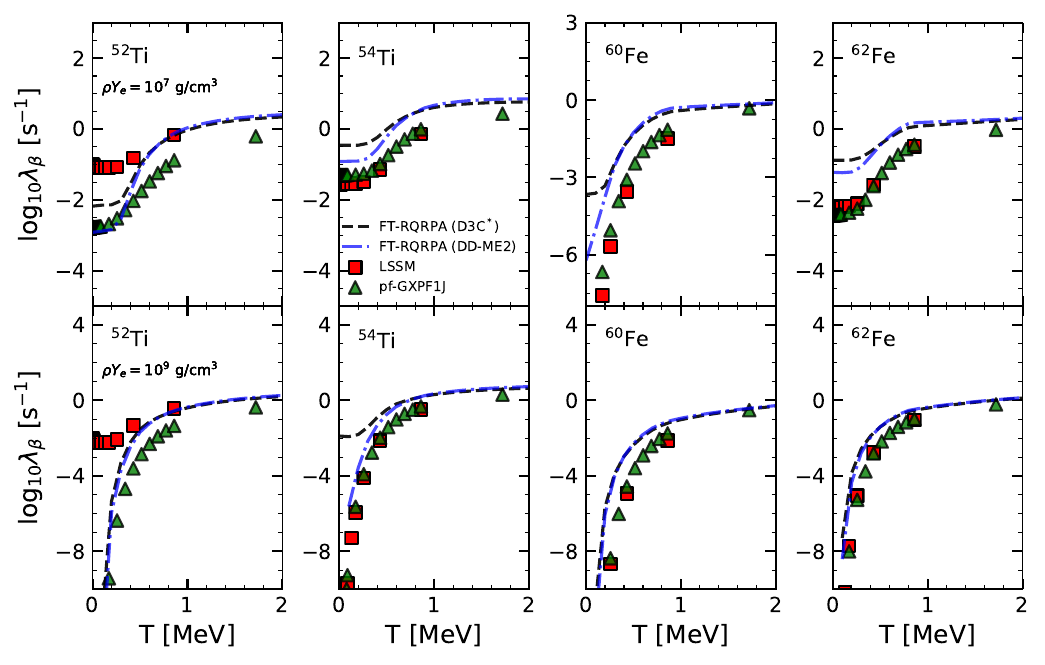}
\caption{Beta-decay rates $\lambda_{\beta}$ for selected nuclei in the temperature range T = $0-2$ MeV for densities $\rho Y_e = 10^7$ g/cm${}^3$ (upper panels) and $\rho Y_e = 10^9$ g/cm${}^3$ (lower panels). The FT-PNRQRPA calculations based on the D3C${}^*$ interaction (black dashed line) and DD-ME2 interaction (blue dash-dotted line) are shown together with the LSSM (red squares) \cite{LANGANKE20011} and shell-model calculations based on the pf-GXPF1J interaction (green triangles) \cite{Mori_2016, suzuki_private}.}\label{fig:comparison_sm}
\end{figure*}

\subsection{Dependence of finite-temperature $\beta-$decay rates on stellar density}

It is also of interest for astrophysical applications to study the dependence of $\beta$-decay rates on the density $\rho Y_e$. Therefore, we also consider the changes in the $\beta$-decay rates at higher stellar densities with increasing temperature. In Fig. \ref{fig:comparison_sm} we compare our results (FT-PNRQRPA) for selected $pf-$shell nuclei calculated using D3C${}^*$ (black dashed line) and DD-ME2 (blue dash-dotted line) interactions in the temperature range T $ = 0-2$ MeV with the large scale shell-model results (LSSM) from Ref. \cite{LANGANKE20011} and shell-model results based on pf-GXPF1J interaction from Refs. \cite{Mori_2016, suzuki_private} at $\rho Y_e = 10^7$ g/cm${}^3$ (upper panels) and $\rho Y_e = 10^9$ g/cm${}^3$ (lower panels). Since the shell-model calculations do not include first-forbidden transitions, we only use the allowed Gamow-Teller transitions in the calculations for the comparison. Comparing the upper and lower panels of Fig. \ref{fig:comparison_sm}, it can be seen that $\beta$-decay rates significantly decrease with increasing density. It is also seen that the FT-PNRQRPA calculations using both D3C${}^*$ and DD-ME2 interactions are in good agreement with both shell-model results, and  overall trends of increasing rates with increasing temperature are well reproduced. In principle, with increasing temperature at fixed $\rho Y_e$, the electron chemical potential $\mu_e$ slightly decreases, allowing more states in the $\beta$-decay energy window. Again, due to increasing contribution of de-excitations, rates continue to increase with temperature also at $\rho Y_e = 10^9$ g/cm${}^3$, having good agreement with shell-model results at finite temperatures (see the lower panels of Fig. \ref{fig:comparison_sm}).
The agreement is better at higher temperatures where individual nuclear properties (e.g. shell structure and pairing) become less important due to the larger number of excited states. It should be noted that at high temperatures inclusion of negative energy transitions becomes very important in obtaining reasonable agreement with shell-model calculations. The differences between the $\beta$-decay rate predictions stem from the assumptions of the models, as expected. Note that in contrast to shell-model calculations, which assume Brink hypothesis to treat transitions from highly-excited states, our model makes no such assumptions. Inclusion of de-excitations follows from equating the FT-(R)QRPA strength function to physical strength function as in Eq. (\ref{eq:phys_equiv}).

Considering these findings, we conclude that the inclusion of negative energy transitions in the calculation of $\beta$-decay half-lives is crucial to obtain a better understanding of the temperature evolution of the half-lives. Especially at lower temperatures, the inclusion of de-excitations in the $\beta$-decay calculations counterbalances the sudden disappearance of the GT${}^-$ strength within the $Q_\beta$ window due to the vanishing of pairing properties around the critical temperature and leads to a smooth decrease of the $\beta$-decay half-lives. Consequently, the FT-PNRQRPA and shell-model results become compatible with each other.

\begin{figure*}
\centering
\includegraphics[width=0.48\linewidth]{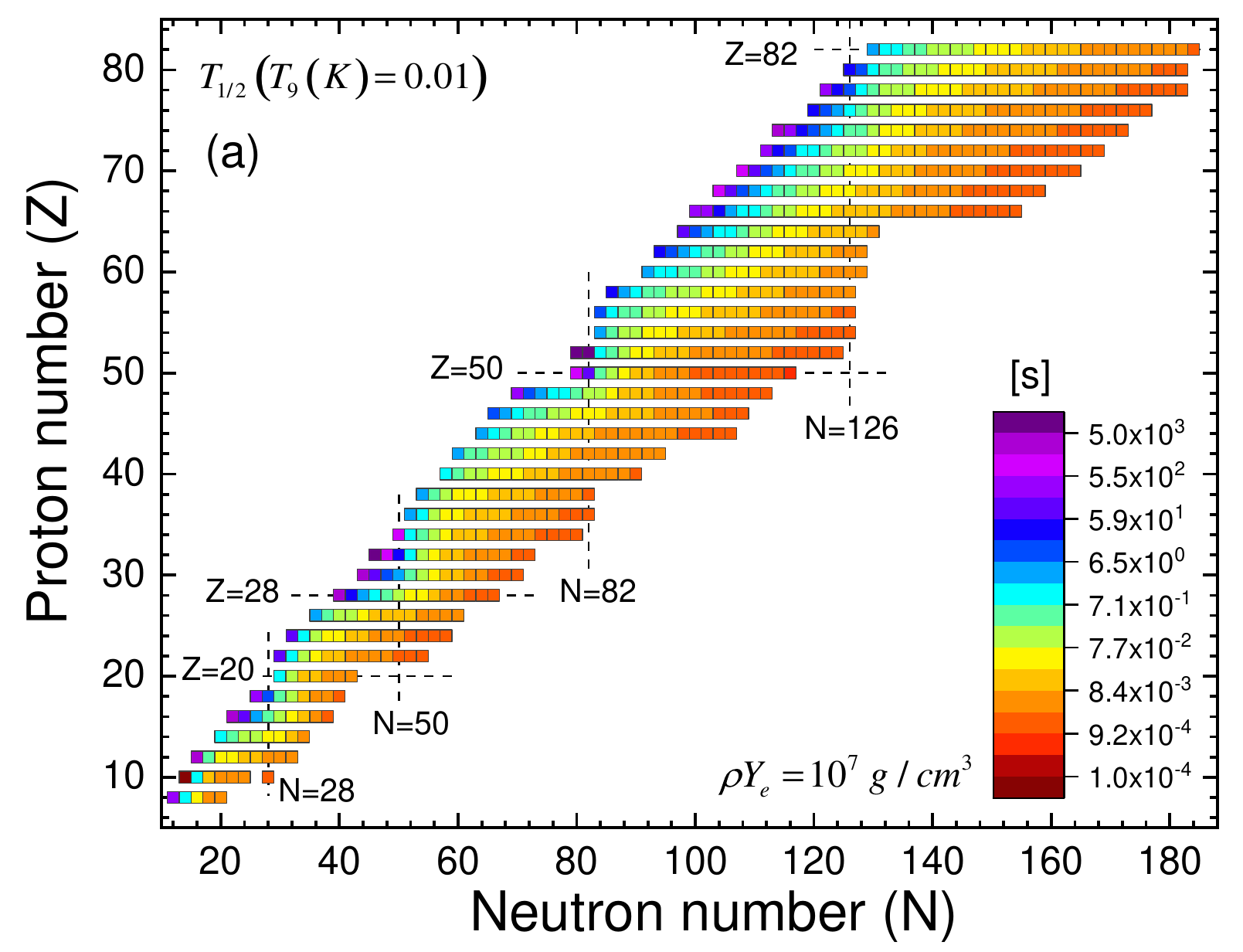} 
\includegraphics[width=0.48\linewidth]{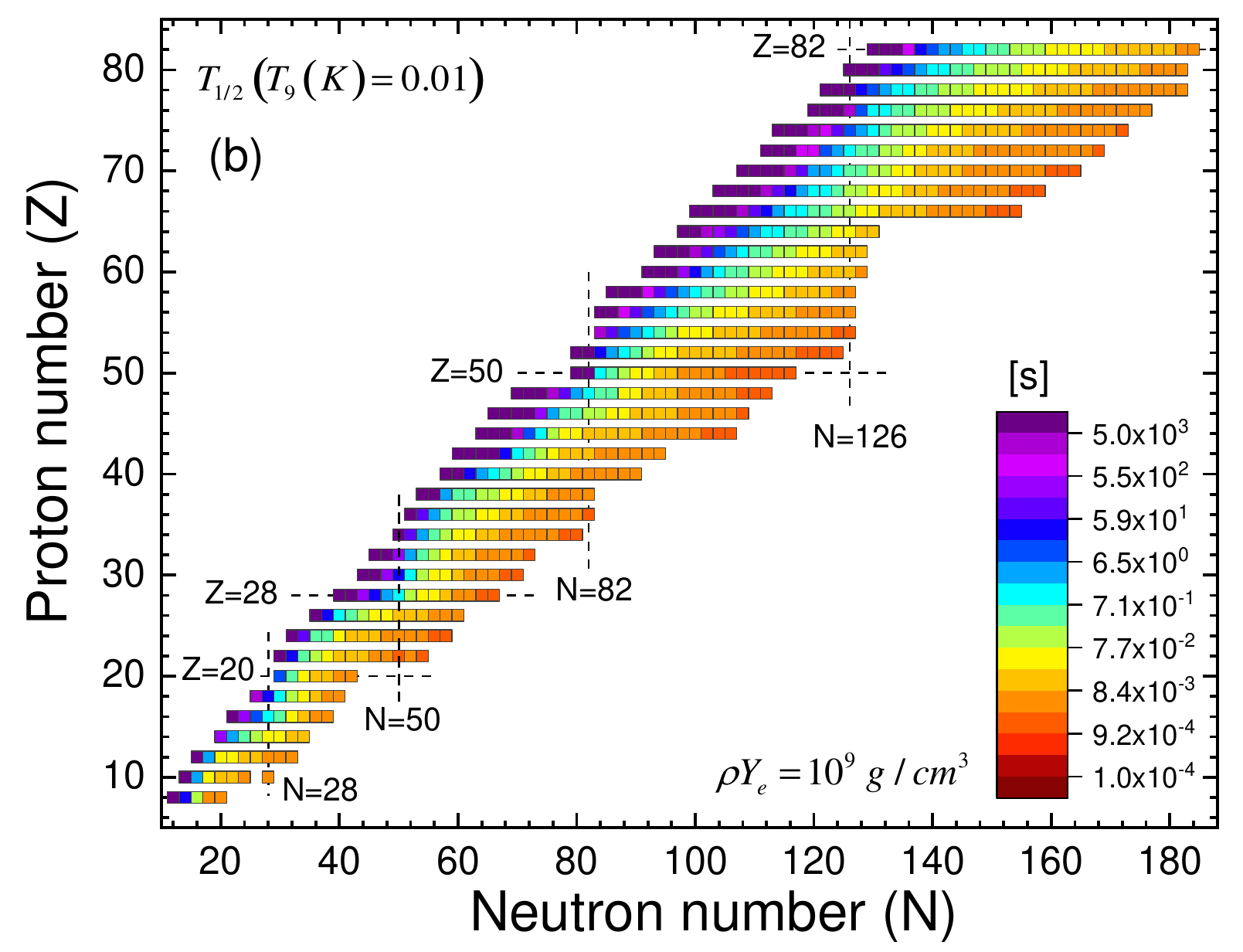} 
\caption{Left panel: the $\beta$-decay half-lives $T_{1/2}$ of nuclei at zero-temperature. Large-scale calculations are performed for even-even nuclei in the range $8 \leq Z \leq 82$, and the stellar density is fixed to  $\rho Y_e = 10^7$ g/cm${}^3$. The even-even nuclei are shown side-by-side for demonstration purposes. Right panel: The same but the density is fixed to $\rho Y_e = 10^9$ g/cm${}^3$.}\label{fig:zero}
\end{figure*}

\begin{figure*}
\centering
\includegraphics[width=0.48\linewidth]{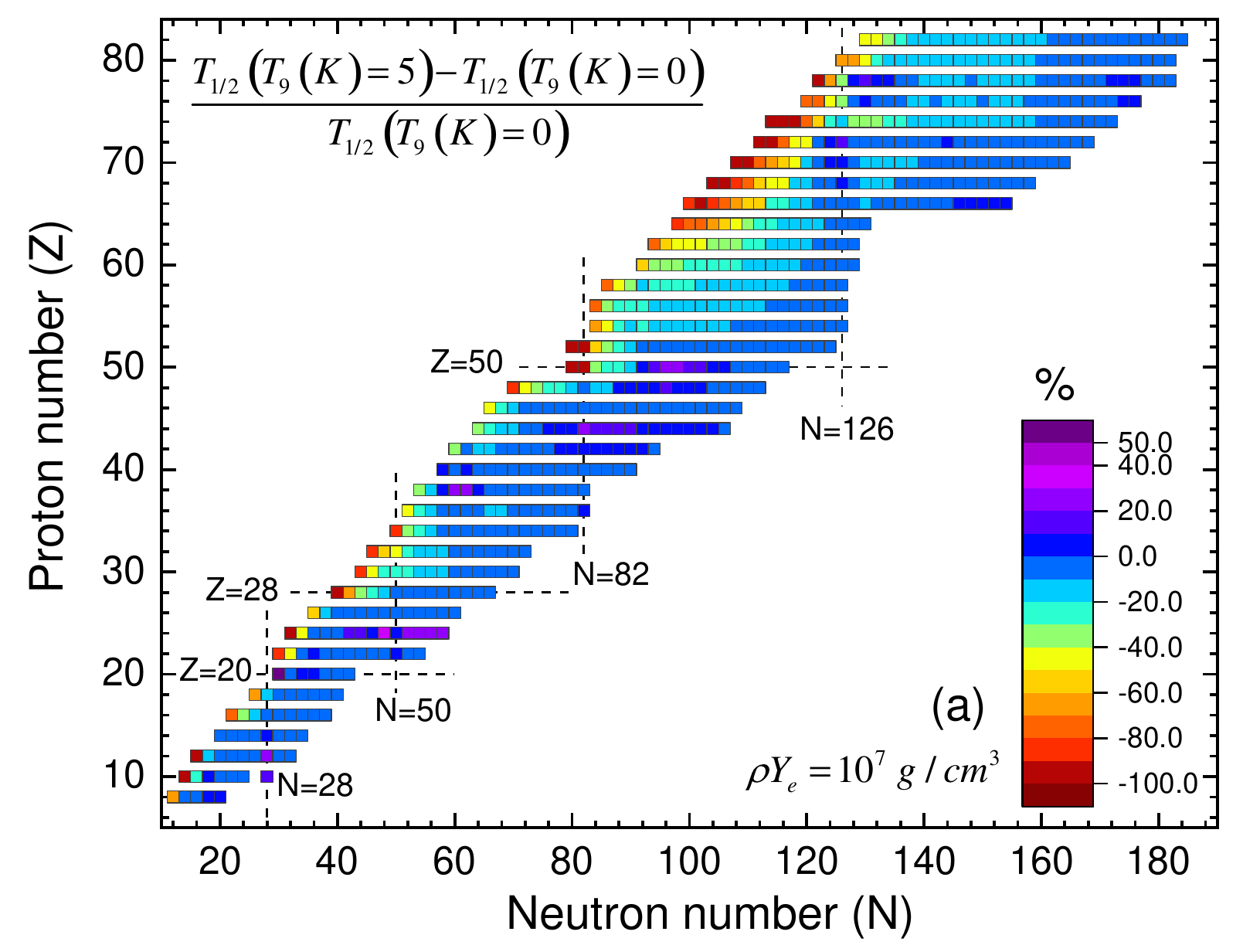} 
\includegraphics[width=0.48\linewidth]{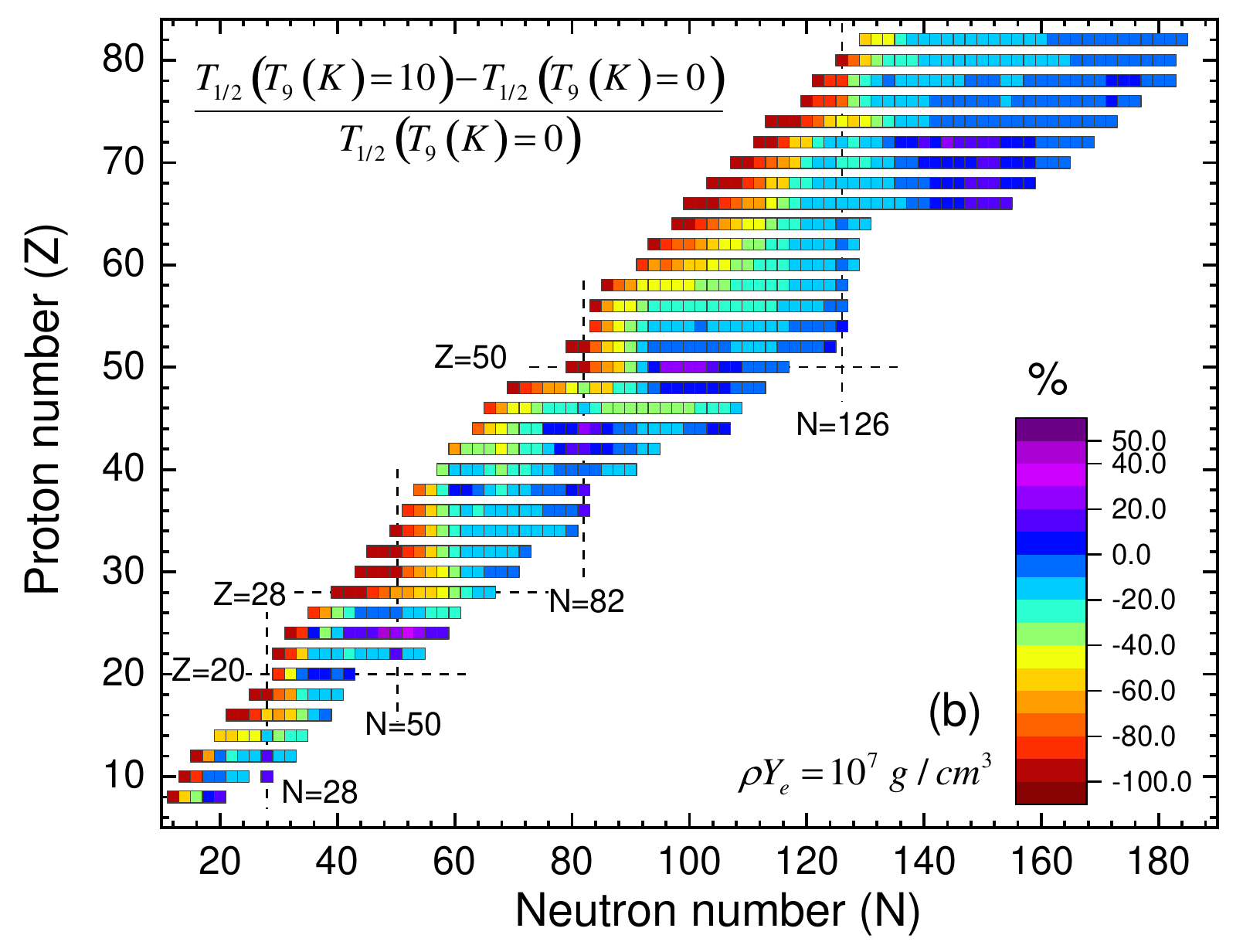} 
\includegraphics[width=0.48\linewidth]{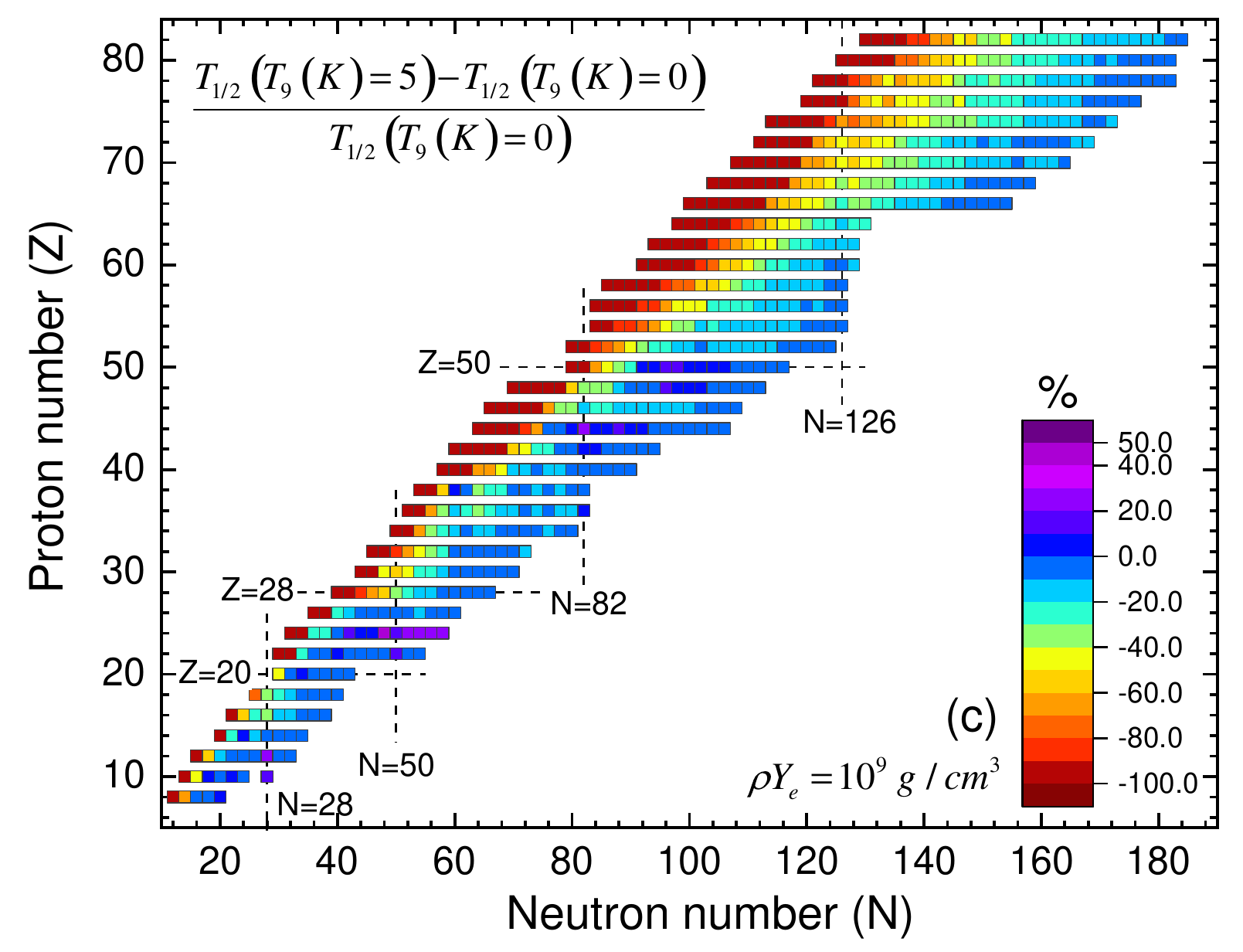} 
\includegraphics[width=0.48\linewidth]{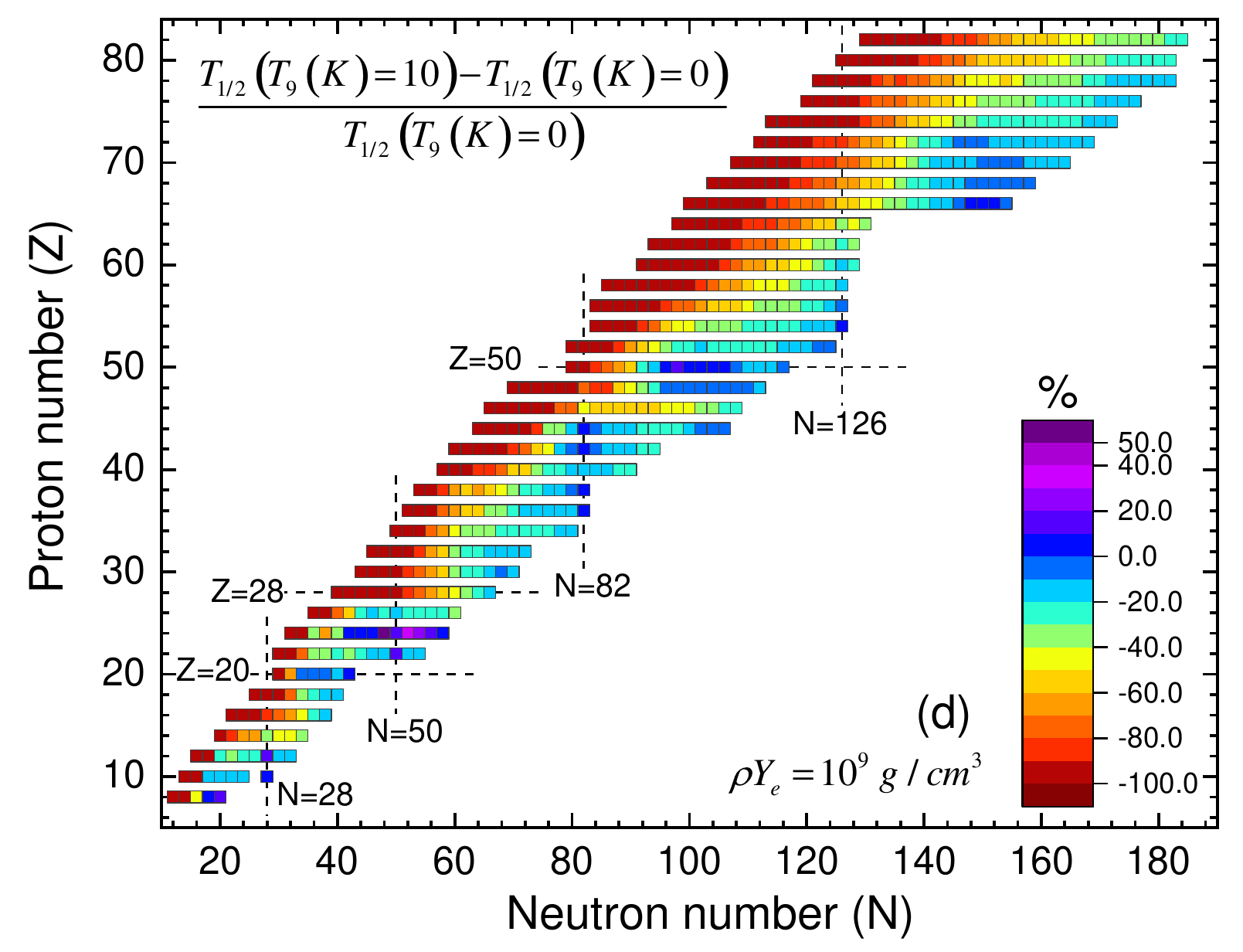} 
\caption{(a)-(b) The percentage change of the $\beta$-decay half-lives $T_{1/2}$ at finite temperatures $T_9(\text{K}) = 5$, $T_9(\text{K}) = 10$ with respect to $T_9(K) = 0.01$ (zero-temperature) results. Large-scale calculations are performed for even-even nuclei in the range $8 \leq Z \leq 82$, and the stellar density is fixed to $\rho Y_e = 10^7$ g/cm${}^3$. The even-even nuclei are shown side-by-side for demonstration purposes. (c)-(d) The same but the density is fixed to $\rho Y_e = 10^9$ g/cm${}^3$.}\label{fig:large_scale}
\end{figure*}

\subsection{Large-scale calculations at zero and finite temperatures}

After investigating the effects of temperature in $\beta$-decay half-lives, we extend our investigation throughout the nuclide chart. In this work, we focus on $\beta$-decay half-lives of even-even nuclei in the $8 \leq Z \leq 82$  range. Following previous work from Ref. \cite{PhysRevC.93.025805}, we perform our calculations for nuclei with half-lives below $10^4$ s at zero-temperature. Pairing gaps for open-shell nuclei are calculated by adjusting monopole pairing constants $G_{n(p)}$ (see Refs. \cite{PhysRevC.101.044305}) to pairing gaps obtained from five-point formula \cite{Bender2000}. Half-lives of doubly-magic ${}^{132}$Sn and ${}^{78}$Ni nuclei are adjusted to the available experimental half-lives by modifying $g^\prime$ coupling of Landau-Migdal term in Eq. (\ref{eq:landau}) as previously described in Sec. \ref{sec:resb}. The same value of $g^\prime$, as determined for the doubly-magic nucleus, is used throughout the rest of Sn and Ni isotopic chain. This is done to compensate for missing strength in doubly-magic nuclei due to ommission of complex-configurations within the (Q)RPA. We present only the results for the bound nuclei with negative chemical potential $\lambda_{n(p)}<0$ \cite{nature_drip_lines}.

In Fig. \ref{fig:zero} we first display the $\beta$-decay half-lives of 705 even-even nuclei at $T_9 (\text{K}) = 0.01$ ($T_9(\text{K})$ denoting temperature in $10^9$ K units) using $(N, Z)$ charts. This is essentially zero-temperature result, however, due to $\rho Y_e$ dependence non-vanishing temperature is usually introduced to keep the calculations finite. In order to demonstrate the effects of the stellar density on the $\beta$-decay half-lives, the calculations are performed at $\rho Y_e = 10^7$ g/cm${}^3$ (Fig.\ref{fig:zero}(a)) and $\rho Y_e = 10^9$ g/cm${}^3$ (Fig.\ref{fig:zero}(b)). It is shown that the $\beta$-decay half-lives decrease with increasing neutron number in both panels, as expected. Increasing the density to $\rho Y_e = 10^9$ g/cm${}^3$, half-lives are predicted to be longer for all nuclei compared to the calculations with $\rho Y_e = 10^7$ g/cm${}^3$. We also notice that the $\beta$-decay half-lives increase considerably in the less neutron-rich side of the $(N, Z)$ chart, and some nuclei become even stable at high densities.

Let us explain the physical mechanism of the changes in the $\beta$-decay rates of nuclei with increasing density. From the rate equation (\ref{eq:beta-decay}), it can be easily deduced that with increasing density ($\rho Y_e$) $\beta$-decay rates decrease rapidly. Because of the increased chemical potential of electrons $\mu_e$ at higher densities, the Fermi-Dirac factor in Eq. (\ref{eq:fdirac}) increases. This factor enters Eq. (\ref{eq:beta-decay}) as $1-f_e(W)$, thus reducing (increasing) the $\beta$-decay rate (half-life) by limiting the number of available excitations in the $\beta$-decay energy window. This is exactly the opposite case of the electron capture, where higher densities excite a larger part of the strength function \cite{PhysRevC.83.045807, PhysRevC.102.065804}. Although densities considered within this work might be too high for the r-process, they could be of significance for the evolution of core-collapse supernovae, especially in the stage when the collapse reaches $A \approx 60$ where $\beta$-decay can compete with the electron capture \cite{Martinez_Pinedo_2000}.

Similar calculations are also performed at finite temperatures, and the effect of the temperature is studied on the $\beta$-decay half-lives of nuclei. To this aim, the calculations are performed at $T_9(\text{K}) = $ 5 and 10, and densities are taken as $\rho Y_e = 10^7$ g/cm${}^3$ and $\rho Y_e = 10^9$ g/cm${}^3$.
In Fig. \ref{fig:large_scale} we display the percentage change in the $\beta$-decay half-lives of nuclei at finite temperatures with respect to the $T_9$(K) = 0.01 (zero temperature) case. Already at $T_9(\text{K}) = 5$ for densities $\rho Y_e = 10^7$ g/cm${}^3$ (Fig. \ref{fig:large_scale}(a)) and  
$\rho Y_e = 10^9$ g/cm${}^3$ (Fig. \ref{fig:large_scale}(c)) temperature effects start to become visible. In the majority of the nuclide map, a decrease in the $\beta$-decay half-lives is obtained with increasing temperature, as expected from previous analysis. Also, some nuclei display a slight increase in half-lives, being located in between the closed shells and in the proximity of neutron drip line. As mentioned above, a decrease in half-lives is mainly related to the changes in the pairing and excitation properties of nuclei with increasing temperature. By inspecting Fig. \ref{fig:large_scale}(a) and Fig. \ref{fig:large_scale}(c), the $\beta$-decay half-lives of nuclei are generally impacted more by the temperature effects for $\rho Y_e = 10^9$ g/cm${}^3$, when compared to the results with $\rho Y_e = 10^7$ g/cm${}^3$. At $T_9(\text{K}) = 5$, the percentage change in the half-lives are found to be below 20\% for both densities near the neutron-drip lines. Going towards the valley of $\beta$-stability, the temperature leads to an important decrease in the half-lives, and the obtained percentage decrease reaches 100\% for the calculations using $\rho Y_e = 10^9$ g/cm${}^3$. It is known that high density leads to an increase in the half-lives of nuclei by decreasing the $Q_\beta$ window at zero temperature. Therefore, the half-lives become more sensitive to the temperature-driven changes in the excitation properties of nuclei.

At $T_9(\text{K}) = 10$ (see Fig.\ref{fig:large_scale} (b) and (d)), half-lives are significantly altered due to the increasing impact of the temperature on the nuclear properties, and many nuclei are showing much larger changes.   Again, nuclei showing the most change are those with initially long half-lives (magic, semi-magic, and close to the valley of stability). Similar to the findings at  $T_9(\text{K}) = 5$,  it can be seen that temperature effects are more pronounced at $\rho Y_e = 10^9$ g/cm${}^3$, changing the half-lives of many nuclei in a considerable way.
We conclude that the  general effect of the temperature is to decrease the half-life of nuclei, especially those in the vicinity of valley of $\beta$-stability. Although some nuclei display an increase in their half-lives with increasing temperature, this effect is mild compared to many other nuclei showing considerable decrease of half-lives. In general, influence of temperature on half-lives depends on the particular shell structure and the pairing properties of nuclei, as discussed in Sec. \ref{sec:resb}.

\section{Conclusion}\label{sec:conclusion}
 
In this work, we have developed a microscopic framework for the description of the temperature dependence of $\beta$-decay half-lives, based on the relativistic nuclear energy density functional with the momentum-dependent meson-nucleon couplings (D3C${}^*$ parameterization). The FT-PNRQRPA has been implemented in the calculations of the allowed and first forbidden transitions for $\beta$-decay, by including both the nuclear pairing and finite temperature effects. The $\beta$-decay properties have been studied by varying the temperature and density, which are relevant for some astrophysical conditions. 

After benchmarking the model to reproduce the measured half-lives at zero temperature, it has been demonstrated that temperature can have a considerable impact on nuclei with longer half-lives, namely, for nuclei with magic numbers and close to the valley of $\beta$-stability. By increasing temperature, the excited states start to shift downward and new states appear in the low-energy region due to the unblocking mechanism of the temperature, which in turn leads to an increase in the $\beta$-decay phase space, and decrease in the half-lives. Furthermore, as the temperature increases, transitions from highly excited states in the parent nucleus i.e. de-excitations, become important in shortening the half-lives around the critical temperatures for pairing phase transition.  Following the example of nuclei in Ti, Fe, Cd, and Sn isotopic chains, we have demonstrated that a significant decrease in half-life occurs near the critical temperature for neutrons. Nuclei with short half-lives at zero-temperature display only a minor effect of temperature. Although those nuclei can also exhibit an increase of half-lives with temperature, on average this effect remains within 10\% of the relative difference with respect to half-life at zero-temperature. It has also been shown that the impact of the forbidden transitions on the half-lives becomes more pronounced with increasing temperature due to the thermal unblocking effect. Increasing the stellar density in the calculations, the $\beta$-decay half-lives increase considerably due to the decrease in the available phase-space.

We have compared our $\beta$-decay rates for ${}^{52,54}$Ti and ${}^{60,62}$Fe with shell-model calculations and obtained reasonable agreement (considering the difference between the models) for both D3C${}^*$ and DD-ME2 interactions, especially at higher temperatures (T $>$ 1 MeV) where $\beta$-decay rates become less dependent on particular shell structure.

The presented model is most suitable for large scale calculations of $\beta$-decay half-lives at finite-temperature throughout the nuclide chart, relevant for astrophysical nucleosynthesis mechanisms. As an initial study toward this direction, the half-lives of 705 even-even nuclei in the range between proton numbers 8 to 82
have been calculated at temperatures $T_9(\text{K}) = 5$ and $T_9(\text{K}) = 10$ and  stellar densities $\rho Y_e = 10^7$ g/cm${}^3$ and $\rho Y_e = 10^9$ g/cm${}^3$.
The strong impact of the temperature and density on the $\beta$-decay half-lives has also been demonstrated over the nuclide map.
Although temperatures where half-lives change significantly appear higher than in some of the nucleosynthesis mechanisms (e.g., $r-$process in Ref.~\cite{MUMPOWER201686}), temperature-dependent $\beta$-decay half-lives could be important in the initial stages of $r-$process \cite{PhysRevC.87.015805} or astrophysical processes like $rp-$process \cite{1981ApJS45389W}, dense thermonuclear explosions and supernovae simulations \cite{JANKA200738} where temperatures are higher. More sophisticated finite-temperature RHB theory, providing accurate scattering of quasiparticle pairs to nuclear continuum, instead of the BCS for the calculation of nuclear properties, is going to be developed and implemented in forthcoming studies. Improved description throughout the nuclide chart by including the half-lives of odd-$A$ nuclei and deformation effects, also necessary for a complete understanding of temperature effects on the $r-$process, are going to be addressed in our future studies.

\section{Acknowledgements}
Discussions with E.M. Ney, J. Engel, R.T. Zegers and S. Giraud about inclusion of de-excitations within the FT-QRPA are gratefully acknowledged. We thank T. Suzuki and M. Honma for their clarification of shell-model calculations. A. R. acknowledges discussions with B. Meyer concerning $r-$process nucleosynthesis. Finally, we thank T. Marketin for sharing his expertise on $\beta-$decay calculations based on relativistic QRPA. This work is supported by the QuantiXLie Centre of Excellence, a project co financed by the Croatian Government and European Union through the European Regional Development Fund, the Competitiveness and Cohesion Operational Programme (KK.01.1.1.01.0004). This article is based upon work from the “ChETEC” COST Action (CA16117), supported by COST (European Cooperation in Science and Technology). We acknowledge support by the US National Science Foundation under Grant PHY-1927130 (AccelNet-WOU:
International Research Network for Nuclear Astrophysics [IReNA]).
Y. F. N. acknowledges the support from the Fundamental Research Funds for the Central Universities under Grant No. Lzujbky-2019-11.

\bibliographystyle{apsrev4-1}
\setcitestyle{numbers,square}
\bibliography{ft_beta_decay_paper}

\end{document}